# Early Prediction and Diagnosis of Retinoblastoma Using Deep Learning Techniques

C. Anand Deva Durai[1], T Jemima Jebaseeli[2], Salem Alelyani[3], Azath Mubharakali[4]

[1]Assistant Professor, College of Computer Science, King Khalid University, Abha, Kingdom of Saudi Arabia.
[2]Assistant Professor, Department of Computer Science and Engineering, Karunya Institute of Technology and Sciences, Coimbatore, Tamilnadu, India.
[3,] Director of Research and Development Center for Artificial Intelligence, College of Computer Science, King Khalid University, Abha, Kingdom of Saudi Arabia.
[4,] Head of Research and Development Center for Artificial Intelligence, College of Computer Science, King Khalid University, Abha, Kingdom of Saudi Arabia.

**Abstract**

Retinoblastoma is the most prominent childhood primary intraocular malignancy that impacts the vision of children and adults worldwide. In contrasting and comparing with adults it is uveal melanoma. It is an aggressive tumor that can fill and destroy the eye and the surrounding structures. Therefore early detection of retinoblastoma in childhood is the key. The major impact of the research is to identify the tumor cells in the retina. Also is to find out the stages of the tumor and its corresponding group. The proposed systems assist the ophthalmologists for accurate prediction and diagnosis of retinoblastoma cancer disease at the earliest. The contribution of the proposed approach is to save the life of infants and the grown-up children from vision impairment. The proposed methodology consists of three phases namely, preprocessing, segmentation, and classification. Initially, the fundus images are preprocessed using the Liner Predictive Decision based Median Filter (LPDMF). It removes the noise introduced in the image due to illumination while capturing or scanning the eye of the patients. The preprocessed images are segmented using the 2.75D Convolutional Neural Network (CNN) to distinguish the foreground tumor cells from the background. The segmented tumor cells are classified and the malignancy of the tumor is classified into different stages and further grouped. The proposed optimization technique improves the algorithm's parameter and suitable for multimodal images captured using a different configurations of disease under different circumstances. The suggested system improves the performance of the proposed approaches' accuracy to 99.82%, sensitivity to 98.96%, and specificity to 99.32%. The proposed approach provides the best solution and an alternative approach for competitive methods.

## 1. Introduction

The retina is the delicate layer that of light covering the back of the eye. It is the reason for vision because when the light signal falls on the eye it will be transmitted to the brain through the retinal nerve fiber layer. Due to retinoblastoma any one of the eyes or both eyeballs gets affected. Hence, the child's eye looks like a cat-eye. Due to the genetic mutation, the retinoblastoma produces the eye cells not to be matured. These cancer cells multiplied spread across the retina. This spread may go to the entire human body; even it affects the spine and brain. The growth of retinoblastoma is increased in the past 60 years and it is identified in one in every 15,000 childbirths. In the US there are 300 cases are identified with retinoblastoma each year [1, 2].

The pathophysiology of retinoblastoma is the genetic part of retinoblastoma. It has laid the foundation of genetics in the anthology itself. Retinoblastoma is one of the very first tumors in the world, where genetics was described and studied. Consider chromosome 13 that is 13q14, where the Rb gene is going to lie. Rb gene is a Tumor Suppressor Gene (TSG). This gene can encode the Rb Protein also it is a Tumor Suppressor Protein (TSP). It suppresses the tumor and control or uncontrolled the proliferation of neoplastic cells. Rb protein is inhibiting the cell cycle at the governor of the cell cycle. It allows the certain proliferation of cells to occur and does not allow the growth of other occurrences [3]. If this governor is compromised and the mutational inactivation of both Rb gene alleles leads to the unchecked development of neoplastic cells, retinoblastoma grows [4]. Hence, retinoblastoma requires both alleles to be inactivated. Depending on the occurrence of inactivation in retinoblastoma, it is classified into two forms as heritable and non-heritable sporadic form [5]. In heritable form, the parent will carry one of the mutant genes to get transmitted through the sperms. It is called the first hit. So the zygote will have one defective gene and all the cells of the child carry one copy of the defective gene and the retinal cells proliferate. Due to this other normal alleles get mutated through these second hits. In retinoblastoma, the first hit already occurred even before the child is born. The second hit happens when the child grows after birth. In non-heritable form, the parent's germ cells, zygotes, and the somatic cells of the child will be normal. Only the first hit occurs in the developmental stage and later, the second hit is acquired. Hence, the two alleles are getting mutated, thereby forming retinoblastoma. In a hereditary tumor, the first hit is inherited during the inherited mutation then, the second hit is acquired through mutation [6]. In sporadic or a non-hereditary tumor, there will be no changes in the parents. But both the hits mutations are acquired after childbirth [7]. These are the two main significant identification of the 2-hit hypothesis.

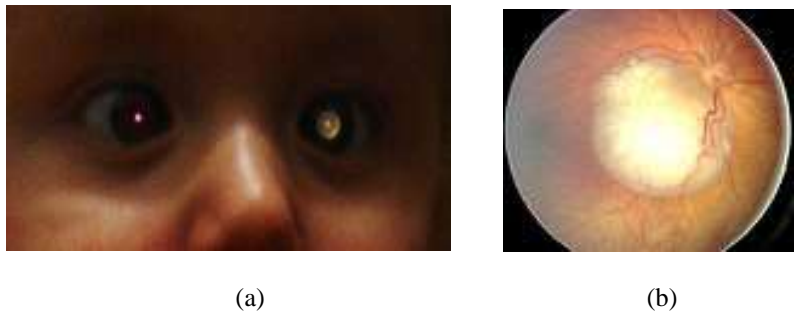

(a)                        (b)

**Figure 1** (a) Child with bilateral retinoblastoma (b) Fundus imaging of the child affected by retinoblastoma.

As shown in Figure 1, the heritable retinoblastoma is always bilateral (B/L) and non-heritable is always unilateral (U/L). Also, in such cases, exceptions may occur. Heritable retinoblastoma occurs in multifocal points in the eye and non-heritable retinoblastoma is unifocal [8]. The patients with a heritable form of retinoblastoma can have secondary tumors like pinealoblastoma called trilateral retinoblastoma. They also include osteosarcoma, sarcoma of the soft tissue, and melanoma. In the case of a non-heritable form of retinoblastoma, the patient won't have chances of getting secondary tumors. In Trilateral retinoblastoma, there will be two blastomas at both eyes. Hence, the patient will have bilateral (B/L) retinal blastoma along with cancer evolving in the pineal gland of pinealoblastoma.

There are some restrictions in the existing segmentation techniques that alter prerequisites, including false malignant identification, triggered by the low separation between the malignant sections and the context. The tumor structure's connectivity deficiency is relatively complicated. The disparity in nanoscopic contrast between the extraction of the object had to be upgraded. In certain approaches, in the event of the malignant fade away from its extended path, it is insufficient to demonstrate the full segmentation effect. Premature termination of the mechanism of tracking results and reduces the entire configuration of the tumor cells. On this motivation, the suggested strategy is based and competently determines the paramount solution of segments identical to the anticipated object. From the pathological images, this model finds the tumor cells. Consequently, small tumors with no discontinuities in their arrangement are qualified to observe.

As follows, the article is organized. The reports and problems related to lesion segmentation approaches are discussed in section 2. Datasets used for experimentation are defined in section 3. Section 4 demonstrates the Liner Predictive Decision-based Median Filter (LPDMF) and 2.75D Convolutional Neural Network's proposed lesion segmentation approaches. The experimental findings and observations are illustrated in Section 5. In Section 6, the paper ends.

## 2. Literature Review

Retinoblastoma grows and proliferates within the eyeball. It arises from the retina in three forms namely, epiphytic, entophytic, and diffuse infiltratively. In entophytic form, the tumor grows into the vitreous space with multiple vitreous seeds and it spread into the exterior subretinal space as subretinal space seeds. Within the eyeball, the tumor fills the vitreous cavity. There it forms subretinal fluid and causes an exudative retinal detachment [9]. Here, the tumor grows behind the subretinal space, and very often there will be the exophytic and entophytic combined growth of tumors at this point of time [10]. The diffuse infiltrative growth mimics the rationalities. It doesn't go behind or forward, but it goes and spreads alone the retina. It is an aggressive growth and grows towards the anterior chamber to form various tumor seeds, and will not have any calcification also they are easily missed.

### 2.1 Histopathology

The pathological retinoblastoma cell consists of small primitive round cells, with necrotic areas and focal zones of dystrophic calcification. The primitive round cells are blue, necrotic areas will be in pink color. The blue nest of cells and a pink sea of necrosis along with calcium deposits of rocks are pathognomonic of retinoblastoma. The three important histopathology features are flexner wintersteiner rosette, homer wright rosette (neuroblastoma), fluorites, and the biopsy is contraindicated. In the flexner wintersteiner rosette, the tumors are growing like a rosette or garland chain. At its center, an empty lumen will be there and looks empty. The cytoplasmic extensions will take place, thereby in the center, and do not have a basement membrane. Whereas, the homer write rosette will have neutrophil in the center of the lumen. The rosettes will indicate the well-differentiated form of tumors [11]. The fluorites are bouquet like structure and which are eosinophilic. It represents the photoreceptor differentiation. The tumor only with laureates is called retinocytoma.

## 2.2 Mode of the spread of retinoblastoma

In the direct spread, the tumor can infiltrate the optic node thereby going into the Central Nervous System (CNS). Also, the tumor spreads in the subarachnoid space and even to the opposite sides of the optic nerve as well. The direct growth may proliferate choroid and sclera in the orbit, thereby it is called orbital retinoblastoma. The hematogenous or blood spread has different meta states to the organs of the liver, bones, and lungs [12]. The trivial spread takes place through the lymphatic to the conjunctiva as well.

## 2.3 Clinical Features

Leukocoria is the white eye reflex that is the most prominent 40 percent manifestation in infants. It occurs with glow, when the eyes are torch lighted, the tumor reflex the white on the pupil. Squinting is the second most common manifestation of retinoblastoma which reduces vision. It occurs at the macular location of the tumor and the child cannot see objects. Also, retinoblastoma may present in the cornea and be wrongly identified most of the time with megalocornea of congenital glaucoma. The diminished vision and the red eye are the cause, too. Inflammatory responses in the anterior portion of the body originate from retinoblastoma, causing severe uveitis or acute red-eye disease. Pseudo-hypopyon stimulates the development of numerous tumor seeds in the anterior eye chamber. Also, it mimics like a masquerader of ulcer or uveitis or puss formation. It leads to diffuse infiltrative of retinoblastoma. Proptosis manifestations occur due to tumor necrosis because of inflammation of lids. In an advanced state, it affects the orbit and is called orbital retinoblastoma and causes, the fungating orbital mass leads to protrusion of the eyeball outside. Buphthalmos of Congenital Glaucoma and Orbital Cellulitis tumor necrosis are the advanced features of retinoblastoma. Buphthalmos are large carnia mimics like retinoblastoma. Cellulitis is orbital retinoblastoma, Diffuse Infiltration is mimic like hypopyon. Therefore a careful diagnosis is required to identify retinoblastoma. Coats disease is similar to retinoblastoma. It has a typical yellowish reflex of xanthocoria whereas, retinoblastoma has a white reflex. The coat is unilateral and often has telangiectasia in the tortuosity of retinal vessels with massive lipid exudates and no calcification.

## 2.4 Segmentation techniques

There has been a considerable amount of research on the development of medical expert systems to automate diagnostic processes [13]. Expert systems that offer straightforward responses based on pre-defined rules; the application of static rules, however, results in minimal understanding, and thus in the lack of a solution to new situations. The emphasis is on moving training data into machine learning through the enhancement of machine learning algorithms. There are various researchers who suggested solutions for identifying retinoblastoma through different studies.

The system introduced by Carlos et al. [14] was able to segment the retinoblastoma cells using CNN. The algorithm doesn't simplify the longitudinal tracking of ocular tumors. Hence, the classification results made the system fail in evaluating the competing treatment strategies. Also, the complexity of the algorithm is increased due to its feature map generated at each level. The algorithm is fully dependent on the expert's annotated data for the classification of tumors. In clinical diagnosis, this may not be a suitable method to do quick analysis since it's difficult to get the expert's data for each real-time clinical image. Also, the intensity of the retinal blood vessels is very less compared to the background. The preprocessing technique fuses some of the blood vessels with the background [15]. The extreme learning machine algorithm segment the images based on the static thresholding and produces oversaturated results and this may not be a suitable technique for clinical tumor classification. Also in the computation of the extreme learning machine model used for generating the hidden layer output, the computational matrix depends on the singular value decomposition. This gives very low efficiency over the higher-order matrix of multimodal images [16]. Huu et al. [17] presented a model using the T1-weighted 3T MRI, where the physiological blinking of the eye incorporates motion artifacts. Also, the hemorrhage tumors and tantalum clips reduced the precision of this model. In the proposed system, to overcome the issues faced in the existing methodologies, well-defined parameters are introduced in the algorithm to improve the accuracy of clinical diagnosis and treatment.

The studies using the retinal fundus image have previously been used to distinguish retinoblastoma by using visual enhancement, segmentation using canny edge detection to differentiate objects from the context. An algorithm called apriori [18] is applied to the resulting pixels. To define retinoblastoma, fundus cameras were then used to process fundus images using the Gaussian filter, rapid Fourier transformation, and then log transformation to compress light pixels in the image [19]. The intense learning method has been developed [20] to resolve the limitations of feed forward artificial neural networks, especially in terms of learning speed.

## 3. Materials

Screening of newborns of retinoblastoma is essential to find early changes. The evaluation consists of analyzing the detailed history of the family. Examination under anesthesia is the thorough indirect ophthalmoscopy key examination. The USG B scan is used to identify the calcification process. It shows the hypertonic stippled dots, which are typical of retinoblastoma. MRI, CT, and fundus methods are used for examination. MRI pictures show the extraocular extension or orbital extension involvement of soft tissue delineated by this scan. MRI scan represents hyperreflective and hyporeflective occurrence of retinoblastoma. CT scan shows the calcified spot and nucleates the eye through the large tumor which is visible. The Retcam pediatric camera is used to capture the fundus image of the patients. The proposed research considers the patients with the age group of 12 to 20 years. The mean times for diagnosis are 0 to 7 days. The identified experts mark the ground truth for evaluation. Also, the public dataset images of the American Society of Retina specialists available in http://imagebank.asrs.org will be used for evaluation. The proposed method considers 243 pathogenic retinoblastoma images for experimentation.

## 4. Methodology

The most important objective of this analysis is to classify each pixel as a lesion or not in a fundus image. A new technique for lesion identification in retinoblastoma images is suggested in this article. It clarifies the novelty of the proposed approach as follows:
  i. During the estimation process itself, the linear prediction filter replaces the noisy pixels quite early on. This improves the similarity between the denoised pixels present in the image.
  ii. The proposed 2.75D CNN collects the details of the hierarchical texture. Due to this function, with fewer parameters, elevated classification efficiency is achieved.
  iii. The generation of patches from three different angles reduces the cost of computing.
  iv. In contrast to 2D and 2.5D CNNs, 2.75D CNN gathers the spatial information that would be useful for the segmentation process.
  v. The accuracy is also improved by 2.75D CNN with fewer parameters compared to 3D CNN.

Figure 2 displays the schematic scheme of the proposed system is shown in Figure 2. Here, the patients have to scan their retina through the fundus camera attached to the iPhone. The captured fundus images are supposed to be loaded into the android application (Rbapp) interfaces. The proposed application is designed with deep learning based image processing algorithms. Initially, the retinal images must undergo a preprocessing through the Liner Predictive Decision based Median Filter (LPDMF). This method improves the contrast and removes the noise present in the retinal image. The retinal green channel has large blood vessels and malignant pixels strength values. This picture is then used for preprocessing on the green channel. The real-time clinical images are needed to be processed by an unsupervised segmentation algorithm. Thus the images are segmented and classified into blood vessels and lesions using a 2.75D CNN based deep learning model. This algorithm is an unsupervised model and contains no training information. All the blood vessels marked by the algorithm are extracted and the obtained vascular map is used for the diagnosis of blood vessels. Retinoblastoma will change the dimension in the blood vessels. The vascular width measurement enables the system to predict the deviation from the normal vessel width structure.

Based on the dimensional of the blood vessels and the number of lesions present in different quadrants of the images are used to find out the severity of the disease. Based on the severity of the disease, retinoblastoma is graded and its stages are identified.

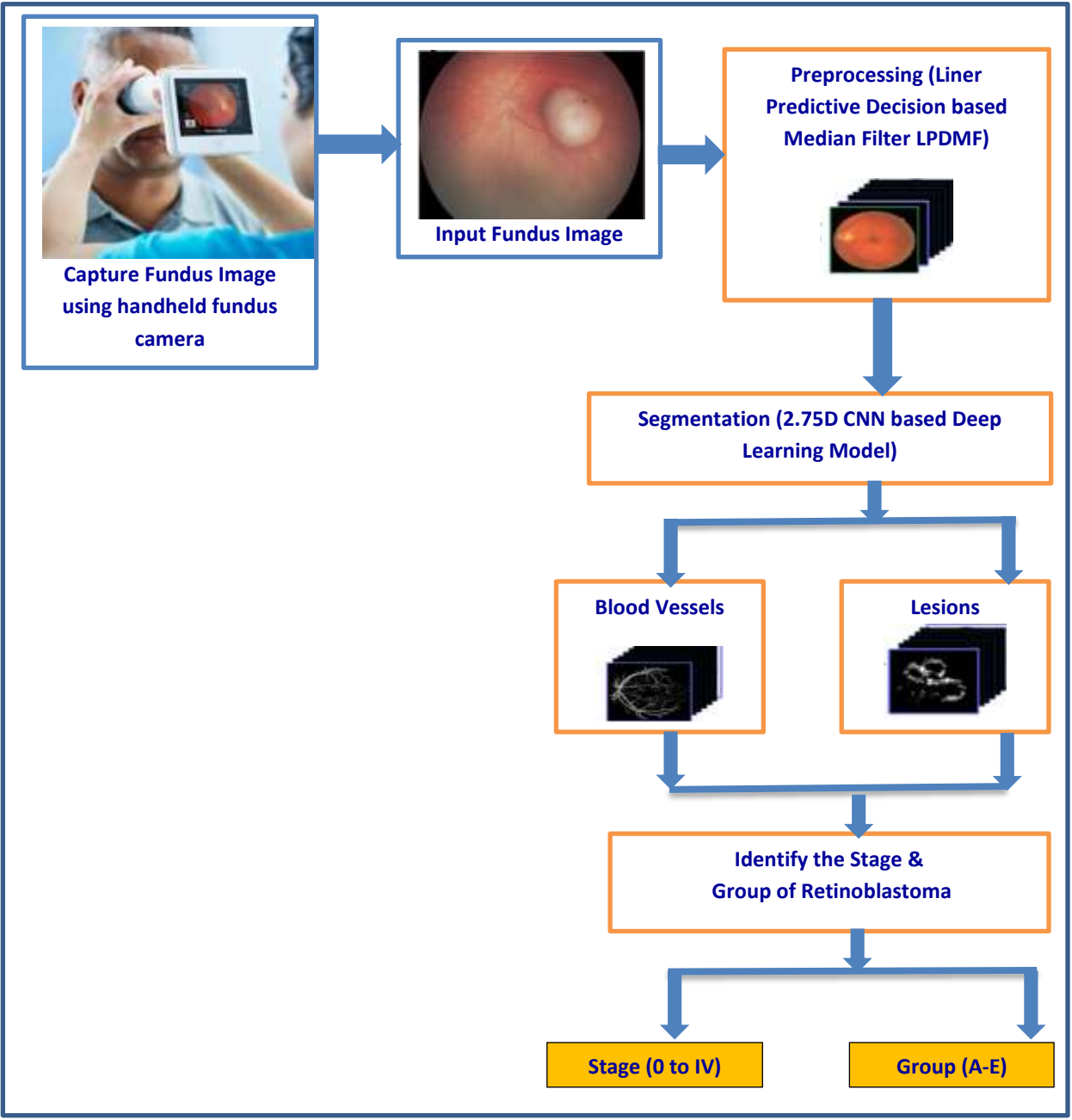

**Figure 2** The proposed architecture for diagnosis of retinoblastoma.

The international retinoblastoma classification of Murphree has been used for grouping the stages. Staging and grouping are the two elements. Staging says the child's survival whereas grouping says the salvage of the eye. There are five stages from 0 to IV. Stage 0 is no enucleated, stage I is enucleated and the tumor can be completely resected, stage II is enucleated but with the microscopic remnants of the tumor, stage III is a regional extension, and stage IV is metastasis. Grouping decides the treatment of retinoblastoma. There are five groups from A to E. The probability of Group A is very low, B is low, C is mild, D is high and E is very high. Group A has a tumor of less than 3mm with no subretinal seeding from the optical disc or fovea. In Group B, the scale of the tumor can be anywhere, but no seed is detected. In group C, the tumor is confined or with local and vitreous seeding. Here, the tumor nipple cell is broken down and spread into the vitreous cavity. In group D, diffuse vitreous seeding and multiple diffused tumor cells spread into the vitreous cavity. Group E is an advanced and extensive stage. In this stage, the tumor is of type diffuse infiltrating and grows as intro ocular hemorrhage type, and when the tumor touches the lens, in the intra segment causes neovascular glaucoma or causing orbital cellulitis.

The American Joint Cancer Committee (AJCC 8) recently gave the retinoblastoma 'H' status. The first retinoblastoma cancer got this 'H' status. 'H' means heritable. Heritable cancer is very dangerous [21]. There is three cornerstones of treatment are save life, salvage globe, and preserve the minimal function of the vision. The following are the four treatment options available. Small cancer can be treated by the focal treatment by providing laser or cryotherapy. The local therapy provides the radiation to the tumor through the radiation shield. The most commonly performed treatment is chemotherapy [22]. If there are diffuse spreads, enucleate therapy is provided to the patient [23].

### 4.1 Preprocessing

It is found that noise is present in retinal images acquired from the fundus camera. This noise is introduced either at the stage of image acquisition or the stage of digital transformation into the original image. Owing to the effect of noise, the tiny tumor sizes in the fundus images are uncertain about their identity. To maximize the optimization effects of segmentation, the collected images from different sources should be standardized by pre-processing steps.

**Liner Predictive Decision based Median Filter (LPDMF)**

The noise-free pixel present in the input image $I_x$ is considered as a set $x$ that consists of the elements $x = \{x_1, x_2, ..., x_n\}$. The median of set $x$ is $x_m$. Consider another set $y$ with noise free images hence, $I_y$ contain $y = \{y_1, y_2, ..., y_n\}$ noise free pixels. The noise present in set $y$ is $y = \{y_{t+1}, y_{t+2}, ..., y_{t+n}\}$, the median value of this set is $y_m$. The elements are arranged in a linear order. Make the third set $z$ replaces the set $y$ hence, the new set $z$ becomes $z = \{y_1, y_2, ..., y_n\}$. The median of set z is $z_m$. The noise present in the z set is $z_{noise} = z_{t+1}, z_{t+2}, ..., z_{t+n}$. These pixels are substituted for the noisy pixels in $y$ set. If there are more than 50% of the elements of $y$ set is noisy then, $\|x_{t+1} - z_m\| < \|x_{t+1} - y_m\|$. If the impulse noise is having a high density value, then the substitute set's median value will be derived from the noise free pixels values presented in the original set. This process enhances the correlation of the denoised pixels present in the image.

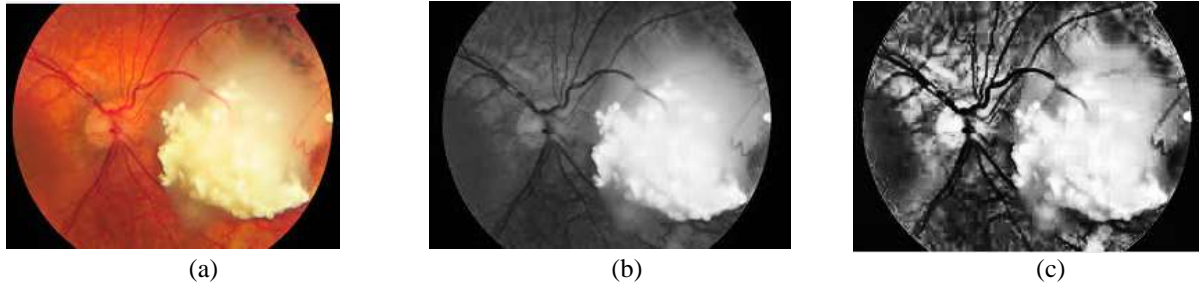

(a)           (b)           (c)

**Figure 3** (a) Input image, (b) Green channel, (c) LPDMF preprocessed image.

Figure 3 shows the input fundus image from the dataset that is considered for analysis. The first step is to extract the green channel of the image. Then, the image is preprocessed by LPDMF that is considered for further segmentation and diagnosis.

### 4.2 Segmentation using 2.75D CNN based Deep Learning Model

2.75D CNN is the multipath segmentation system based on voxel-by-voxel classification, where voxel represents the value of the pixel on a 3D space. There are 2,120 sample points were extracted from a single image. The samples are collected from each orthogonal direction of the plane. As shown in Figure 4, the three dimensions of the input image is given into the system to generate the patch. The generated 2D patches have undergone the process of a 2.75D convolutional neural network. The segmented images of the patches are merged using the bayesian aggregation function to produce the final segmented result.

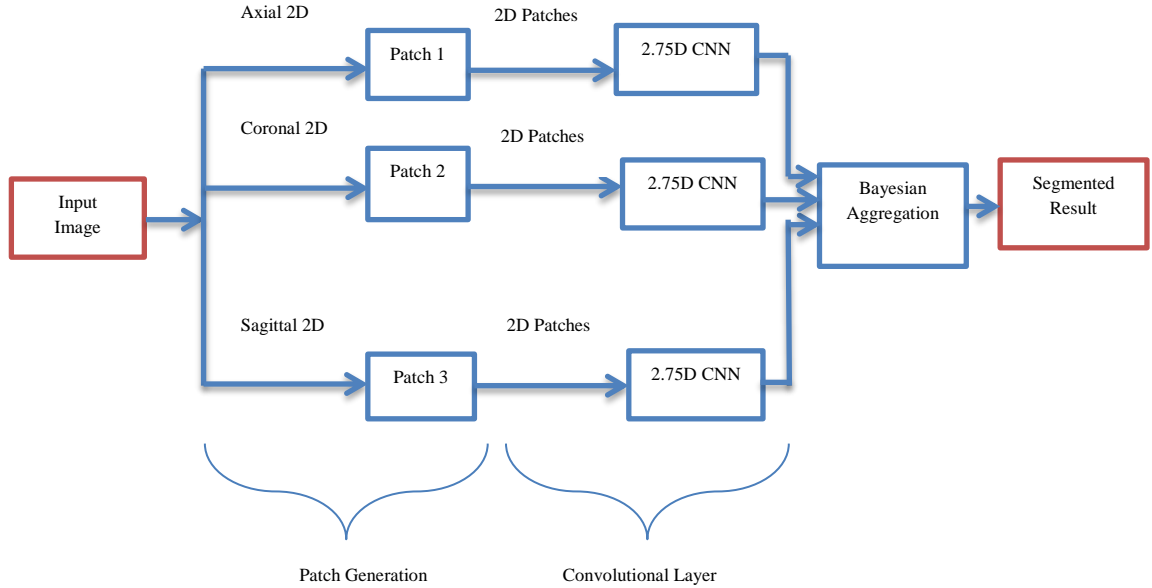

**Figure 4** Architecture of proposed 2.75DCNN model for segmentation of retinoblastoma.

In the proposed model, the patch size is fixed as $64 \times 64 \times 64$ pixels that cover the complete information of the image slices. The orthogonal directions of the 2.75D CNN stack cover axial, sagittal, and coronal planes. The clinical details about the lesions have been collected from all three angles respectively. Hence, the information is collected from three adjacent slices of the image. The proposed model has been trained using 32 patients. During the training process, each patch contains nine slices of the sample. The sampling grid (GR) has three samples at the coordinate direction concerning to x,y,z is $GR^1_{xy}, GR^1_{yz}, GR^1_{xz}$. The size of the samples is $N \times N$ that is formed perpendicular to X, Y, and Z-axis. The patches are rotated, hence the grid is tilted to -45° to 45° and it generates $GR^2_{xy}, GR^2_{yz}, GR^2_{xz}$ and $GR^3_{xy}, GR^3_{yz}, GR^3_{xz}$ respectively. The generated 2.75D patch is $E(GR_i)$ where, $E$ is the grid interpolation of the slice. $GR_i$ is the grid that is corresponding to the $i^{th}$ patch slice.

The proposed scheme applies a novel multi-channel input scheme in 2.75D. It processes the spatial features of pixels similar to 3D CNN while still sampling pixels on a 2D plane. This reduces the algorithm's computation costs. 2.75D CNN is having advantages over 2D and 2.5D due to the requirements of fewer training input parameters and fewer training samples to produce high accuracy. While extracting information through those methods, there were few information losses. Hence to improve the classification accuracy, 2.75D is proposed. Here, the spatial texture information from too many patches from each slice increases the parallel CNN processes.

**Step 1: 2D Patch Extraction**

While transforming a 3D image into a 2D image, the set of sample points on the surface that originates from the north pole ends at the south pole. Hence, the radial lines are formed and the sphere center will be connected. As shown in Figure 5, the CNN extracts the features of multiple patches from the input image. The aggregation function combines all output of CNN. The same weight has been applied over all patches to make CNN compare the features of the patches. The radial line samples are having length L. These radial points are identified by the latitude angle $[0, 2\pi]$ and longitude angle $[0, \pi]$. If the spiral line is around the sphere, the sums of circumferences of several horizontal circles are equal to the length of the spiral line.

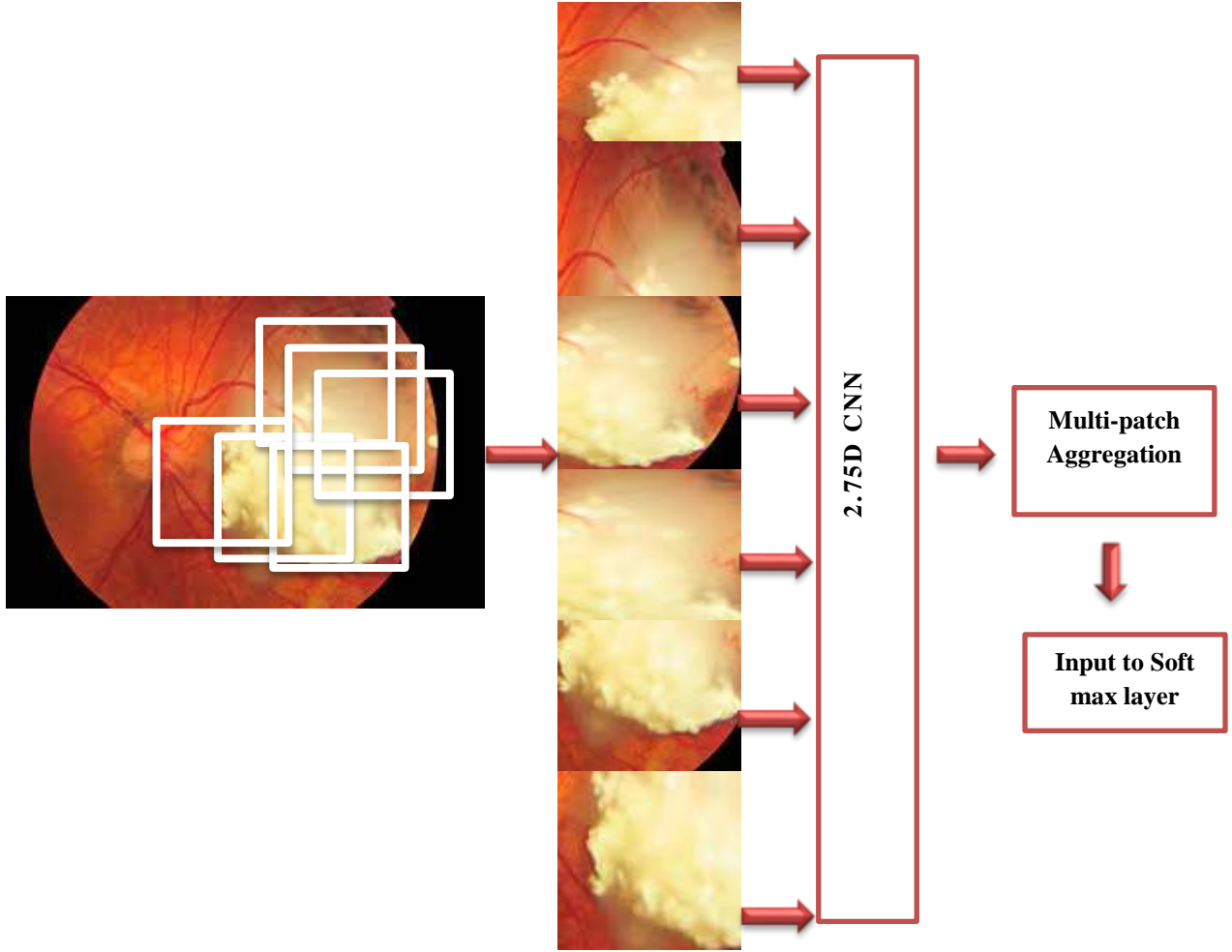

**Figure 5** Multi-patch Bayesian aggregation network.

The total number of surface points on the sphere surface is calculated by the following equation.

$$SP = \sum_{\alpha=0}^{n} 2N\|\sin(\alpha\pi/n)\| \tag{1}$$

$$= 2N \int_0^n \|\sin(e\pi/n)\| d\alpha \tag{2}$$

$$= 2N \int_0^n (n/\pi) \|\sin(e\pi/n)\| / d(\alpha\pi/n) \tag{3}$$

$$= 2N^2/\pi \int_0^\pi \|\sin(\alpha)\| d\alpha \tag{4}$$

$$= 4N^2/\pi \tag{5}$$

Here, N and 2N are the elevation angles which are divided into two sections. $\alpha$ represents the sample points of the horizontal circle. $4N^2/\pi$ gives the number of sample points on the surface. The input image contains $N$ set of patches. To extract the patches from individual images, the features and activation function of the image is used. These features are predicted using the aggregate function.

**Step 2: 2.75D CNN**

2.75D CNN method extracts the features more efficiently for representation in 3D compared to 2D and 3D CNNs. The obtained 64x64x64 nodule volume of the image is considered for evaluation of the proposed 2.75D CNN algorithm.

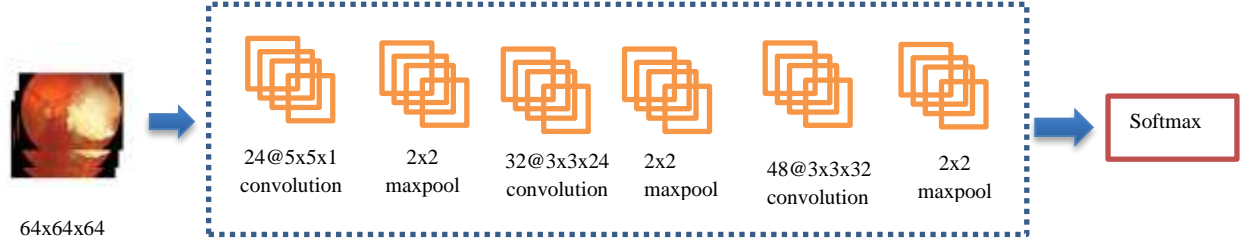

**Figure 6** The proposed 2.75D CNN Architecture.

As shown in Figure 6, the first module is composed of 24 5x5x1 kernels. The second layer is composed of 32 kernels of size 3x3x24. The third convolution layer is having 48 3x3x32-sized kernels. In the pooling layer it down-samples the patch volume by half from 24@60x60 to 24@30x30. Just after patches max-pooling is used where the maximum values in non-overlapping windows of size 2x2. A fully-connected layer is the last layer that produces 16 output units. The activation functions including its Rectified Linear Units (ReLU) on each convolutional layer.

**Step 3: Parameters calculation for CNN**

The CNN parameters are calculated for every layer. The parameters are described as per the following layer-wise. The values of each parameter are given in Table 1.

- Input Layer
- Convolution Layer 1
- Pool Layer 1
- Convolution Layer 2
- Pool Layer 2
- Convolution Layer 3
- Pool Layer 3
- Fully Connected Layer 3
- Fully Connected Layer 4
- Softmax Layer

**Input Layer**

The input layer shape is $64 \times 64 \times 64$ and activation size of this layer is $64 \times 64 \times 64 = 262144$. The input layer is not part of learning and its parameter=0. Because it does not having the learnable parameter.

**Convolution Layer 1**

It is the layer where CNN starts it learning process. To calculate the learnable parameter, the following details such as width $(w)$, height $(h)$, previous layer's filter $(lf)$, and the current layer filter $(cf)$ are used. The CNN layer parameters are calculated as $((w \times h \times lf) + 1) \times cf$. Here, the filter size is 5x5 and stride=1, thus the convolution layer learning parameter is $((5 \times 5 \times 1)+1) \times 8 = 208$. The first convolution layer has 24 kernels, hence the activation shape value is $24 \times 24 \times 8 = 4608$.

**Maxpool Layer 1, 2 & 3**

The max pool layer has no learning parameters. Thus it learns from its specific value due to no backpropagation learning process.

**Convolution Layer 2**

Convolution layer 2 has a filter shape $3 \times 3$, stride=1. Hence the training parameter is $((3 \times 3 \times 24)+1) \times 16 = 3456$. This layer have 32 kernels, hence the activation shape value is $32 \times 32 \times 8 = 8192$.

**Convolution Layer 3**

Convolution layer 3 has a filter shape $3 \times 3$, stride=1. Hence the training parameter is $((3 \times 3 \times 32)+1) \times 32 = 9216$. This layer has 48 kernels, hence the activation shape value is $48 \times 48 \times 8 = 18432$.

**Softmax Layer**

The softmax layer has the current layer filter $(cf) \times$ previous layer filter $(pf) + 1 \times$ current layer filter $(cf)$ as the parameter value.

$pf = 70$
$cf = 10$
$Parameter\ value = (10 \times 70) + (1 \times 10)$
$= 710$

**Fully connected layer 3**

The parameter value of fully connected layer3 is calculated as follows.

$pf = 70$
$cf = 110$
$Parameter\ value = (110 \times 70) + (1 \times 110)$
$= 7810$

**Fully connected layer 4**

The parameter value of fully connected layer4 is calculated as follows.
$pf = 110$
$cf = 70$
$Parameter\ value = (70 \times 110) + (1 \times 70)$
$= 7770$

Table 1 Calculation of the number of parameters in the proposed CNN model.

| S. No. | CNN Layers | Activation shape | Activation size | Parameter value |
|---|---|---|---|---|
| 1 | Input | $64 \times 64 \times 64$ | 262144 | 0 |
| 2 | Convolution Layer 1 | (24,24,8) | 4608 | 208 |
| 3 | Maxpool Layer 1 | (2,2,8) | 32 | 0 |
| 4 | Convolution Layer 2 | (32,32,8) | 8192 | 3456 |
| 5 | Maxpool Layer 2 | (2,2,8) | 32 | 0 |
| 6 | Convolution Layer 3 | (48,48,8) | 18432 | 9216 |
| 7 | Maxpool Layer 3 | (2,2,8) | 32 | 0 |
| 8 | Fully connected Layer 3 | (110,1) | 110 | 7810 |
| 9 | Fully connected Layer 4 | (70,1) | 70 | 7770 |
| 10 | Softmax Layer | (10,1) | 10 | 710 |

**Step 4: Aggregation Layer**

To aggregate the output of CNN labels, the majority voting techniques are applied over the images. Hence, the aggregation function $\mu$ is defined in eqn.(6). Consider $\alpha$ is the sensitivity and $\beta$ is the specificity function to measure the performance of the aggregation of the labeled patches.

$$\mu = \begin{cases} 1 & \alpha > 0.5 \\ 0 & \alpha < 0.5 \end{cases} \qquad (6)$$

$$\alpha = \frac{1}{|N|} \sum_{i=0}^{N} x^i \qquad (7)$$

$N \Rightarrow$ Individual surface points of the sample patches

$x \Rightarrow$ The input of the image patch to the network

$i = \{1,...,N\}$

The input of the image patch $\{x_i, \ 1,...,N\}$ is given to the network. For each annotation of $i \in N$, the classifier is trained using the softmax classifier. The aggregation label is computed from the Bayesian network which is given in eqn.(8).

$$\mu_i = \frac{\alpha_i x_i}{\alpha_i x_i + \beta_i (1 - x_i)} \qquad (8)$$

The final output of the softmax layer gives the segmented result that receives the aggregation of all retinoblastoma patches from the Bayesian network.

**5. Experimental Analysis and Discussions**

The proposed model is implemented using Python. The experiments are evaluated on a GNU/LINUX machine and the algorithm also needs less computation time. Compared with the ground truth image, the structures of the acquired tumor retinoblastoma indicate that the proposed technique is more effective. The overlapping tumors are categorized and correctly traced to the source node. The tumor structure formation is close to the structure of the image of ground truth. The experimental result is shown in Figure 7. In the first iteration, the optical disc is removed to track the malignant accurately from its root node without any misguidance. In the first iteration, the optic disc is separated to prevent misguidance along the shortest path of the tumor cells within the optic disc. The tumor cells detected are connected to their closest neighbors in the next iteration. Within three iterations of the 2.75D CNN model, the entire image is segmented. The morphological process occurs at the top of the image and is scattered throughout the remaining portion. The network transmits are the pixels attached to the next neighbor's pixels. This breaks the boundaries from the foreground pixels which do not interact with the pixels at the end of the rows.

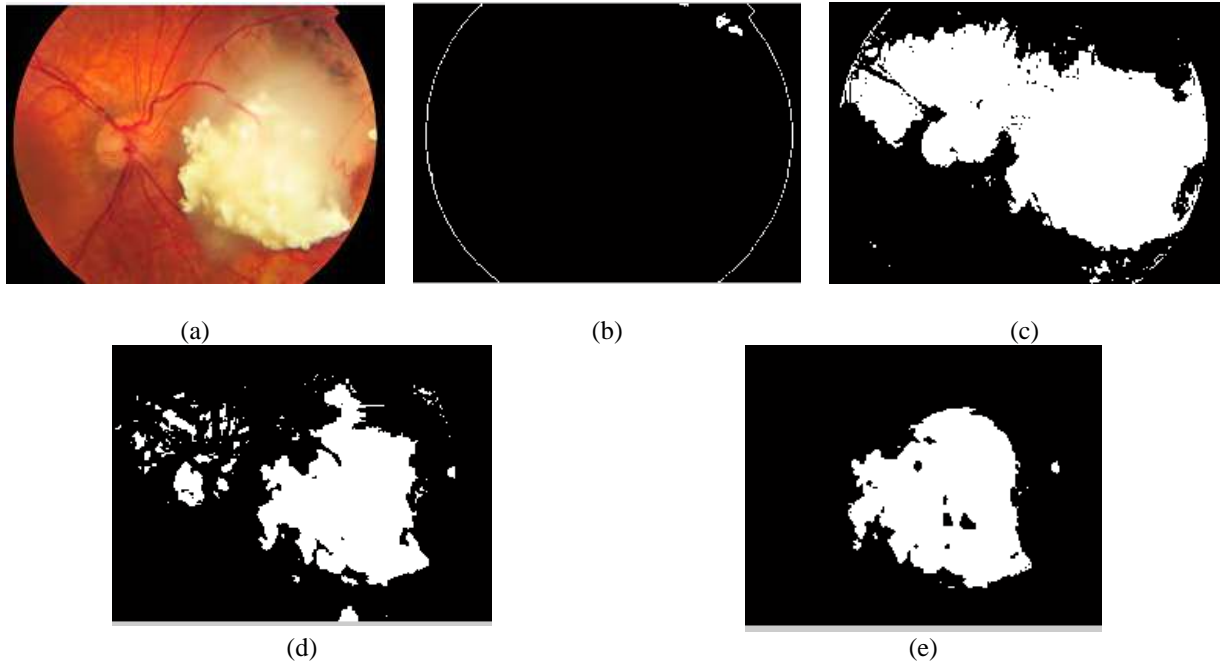

(a)      (b)      (c)

(d)      (e)

**Figure 7** (a) Source Image, (b) Generated Mask, (c-e) Segmentation at iteration-1, iteration-2, Iteration-3, (e) Final segmented result.

The performance outcomes are calculated by the True Positive (TP), True Negative (TN), False Positive(FP), False Negative (FN), Sensitivity, Specificity, and Accuracy parameters. The ground truth of G(i,j) is contrasted to the segmented output image Y in the computation of TP, TN, FP, FN (i,j). The row and column vectors of the respective image are I and 'j.' The same position in pixels of both the image of the ground truth and the segmented image does not contain any tumor pixels, then TN is 1.

$$If(G(i,j)==0 \ \&\&\ Y(i,j)==0)$$
$$TN = TN + 1$$
$$end$$

The tumor is in the same pixel spot as the representation of the ground truth then TP is 1.

$$If(G(i,j)==1 \ \&\&\ Y(i,j)==1)$$
$$TP = TP + 1$$
$$end$$

If the pixel position of the image is where the tumor pixel is present and wrongly marked in the segmented image to be a non-tumor pixel, FP will be 1.

$If(G(i,j)==1 \&\& Y(i,j)==0)$
$FP = FP + 1$
end

If the position of the pixel of a ground truth image is the location of the non-tumor pixel, wrongly defined in the segmented image as the tumor pixel, so FN is 1.

$If(G(i,j)==0 \&\& Y(i,j)==1)$
$FN = FN + 1$
end

$$TPR = Sensitivity \qquad (9)$$

$$FPR = 1 - Specificity \qquad (10)$$

Sensitivity implies the power of the algorithm to detect the pixels of the tumor correctly.

$$Sensitivity = \frac{TP}{TP+FN} \qquad (11)$$

The Specificity of the algorithm is the capacity of the non-tumor pixel to better detect it.

$$Specificity = \frac{TN}{TN+FP} \qquad (12)$$

Accuracy is the degree of proximity of a quantity to the true value of that quantity.

$$Accuracy = \frac{TP+TN}{(TP+FN)+(TN+FP)} \qquad (13)$$

To distinguish a single malignant with no variation, the proposed technique is adequately comprehensive. To find non-malignant pixels, the algorithm is sensitive. The grouping of tumor pixels from others is also competent. The performance of the proposed method, as given in Table 2, indicates an average of 98.96% sensitivity, 99.32 % specificity, and 99.82% accuracy in the normal and abnormal pathological images.

**Table 2** Performance comparison of segmentation of retinoblastoma.

|  | Sensitivity | Specificity | Accuracy |
|---|---|---|---|
| **2D** | 70.35 | 94.69 | 96.24 |
| **2.5D** | 72.9 | 96.12 | 97.31 |
| **2.75D** | 98.96 | 99.32 | 99.82 |

The performance of the proposed method is calculated using the Receiver Operating Characteristic (ROC). ROC is a TPR relative to the FPR map by changing the limit. The ROC includes a schematic value of the TPR and FPR axes for the y-axis. In Figure 8, the ROC for both normal and malignant images is shown. On the ROC curve, the suggested approach is defined. The values are equal to 99.68%, which indicates that the efficiency of the technique suggested has been enhanced. ROCs are 89.54% for 2D and 91.2% for 2.5D CNN techniques.

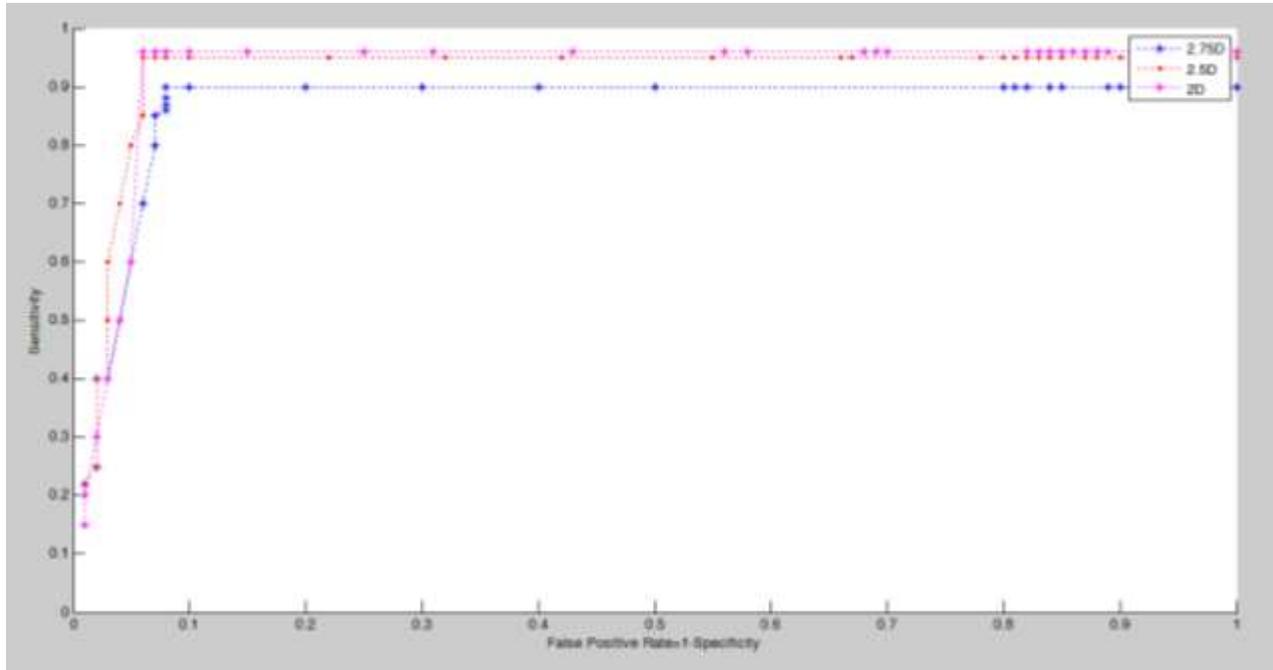

**Figure 8** ROC of segmented retinoblastoma images using 2D, 2.5D, and 2.75D CNNs.

Although maximum efficiency has been reached by the proposed approach, few problems remain to be solved. Small tumor pixels are somewhat translated from their ground-truth representation in the end product. Therefore, make use of numerous professionals who manually segment the images of ground truth to be made. There will be greater confidence around the smallest malignant positions and magnitudes in such a venue. The proposed algorithm is not disrupted by light or dark spots in pathological images. Alternatively, a wider collection of classifiers may be used to train the system. To make it independent, an automated process for OD detection should be applied.

Few automatic approaches have been tailored for the segmentation of MRI ocular tumors. The first experiments were devoted to the segmentation of retinoblastoma in infants. A 3D-Unet [24] and a 3D-CNN [25], focused on very small datasets of 16 and 32 eyes of retinoblastoma, were suggested for two deep learning techniques. However, the tumor segmentation performance reported in these pioneering research ventures was relatively low, with an average Dice Similarity Coefficient (DSC) measurement of about 62%. Retinoblastoma tumors have recently been tackled based on image registration and the MRI threshold [26], but only four cases have been qualitatively assessed.

According to the group-wise treatments options are chosen. Group A needs focal therapy, Group B needs Chemotherapy, Group C needs Chemotherapy, Group D needs Chemotherapy, and Group E needs enucleation. The focal therapy is provided through three means such as cryotherapy, photocoagulation, and thermotherapy. In cryotherapy, the tumors are freeze in the peripheral. In photocoagulation, the lasers are given around the tumor, it causes more scar and reduces vision. This may cause tunnel vision or severe macula damage if the tumor is on the macula. More recently, Thermotherapy is the most preferred one. It is called Trans pupillary Thermo Therapy (TTT) where the intraretinal temperature of the tumor is raised. It causes apoptosis of the tumor cells. So laser is on the tumor not around the tumor. So less scar and defect less vision on the patients has occurred. The local therapy consists of External Beam Radio Therapy (EBRT) where radians have given on the tumor locally; Plague Bracy Therapy was a shield of the radioactive shield is placed and sutured to episclera. The shield materials are Iodine-125 and Tuthenium-106 to destruct the tumor continuously.

Chemotherapy is given when there are any intraocular retinoblastoma and extraocular retinoblastoma with the local regional spread. If the tumor is not healed by the focal or local therapy, chemotherapy is given and gives a very successful rate of treatment. Chemotherapy is given most commonly through intra-venous route or intra-arterial or intra-vitreal route. While applying this therapy shrinks the tumor size. Enucleation is the oldest way to save the life of the statement of Group E conditions. Here, the eyeball and the optic nerve as long as possible are going to be removed. So that there will be no spread of tumors. 19-20 mm stump the optic nerve is cut and removed in this method when the centers are having no other therapy. This technique is called as Myoconjunctival technique. Here, the muscle tenons, the conjunctiva is intact and the optic nerve and the eye are removed. Also, an artificial implant is placed and it is sutured with muscle, tenons, and conjunctiva. The conformer is placed to maintain the spaces.

## 6. Conclusion

It is important to correctly identify the malignant in clinical investigation. Different methods exist that do not locate retinoblastoma tumors inside the optic disc of the fundus images. In agreement with the review of outcomes, the success metrics of the proposed methodology are improved. The proposed solution qualitatively segments the tumors without containing any background noise. The findings indicate that the approaches suggested are an appropriate complement to the supervised methods. Also, the experts analyzed the proposed segmentation results and concluded that the quality is not compromised. In diagnosing retinoblastoma patients, the proposed method reduces the task of eye doctors by examining the retina.

**Acknowledgements**

The authors would like to thank Dr. J. Kishore Kumar Jacob, Eye Specialist, K.K. Medical College Hospital, Nagercoil, India, and Dr. H. Hector, Consultant Ophthalmologist, C.S.I. Hospital, Neyyoor, India, for their immense assistance in providing the fundus images of the clinical fund and assessing the findings of the segmentation.

The research is done in the Center for Artificial Intelligence with the financial support by the Deanship of Scientific Research at King Khalid University Under research grant number (R.G.1/210/42).